# #FoundThem - 21st Century Pre-Search and Post-Detection SETI Protocols for Social and Digital Media

**Authors:** Duncan Forgan[1] and Alexander Scholz[1]

1: SUPA, School of Physics and Astronomy, University of St Andrews

Corresponding Author: Duncan Forgan (dhf3@st-andrews.ac.uk)

## Abstract

The transmission of news stories in global culture has changed fundamentally in the last three decades.  The general public are alerted to breaking stories on increasingly rapid timescales, and the discussion/distortion of facts by writers, bloggers, commenters and Internet users can also be extremely fast. The narrative of a news item no longer belongs to a small cadre of conventional media outlets, but is instead synthesised to some level by the public as they select where and how they consume news.

The IAA Search for Extraterrestrial Intelligence (SETI) post-detection protocols, initially drafted in 1989 and updated in 2010, were written to guide SETI scientists in the event of detecting evidence of extraterrestrial intelligence, but do not give guidance as to how scientists should prepare to navigate this media maelstrom.  The protocols assume communication channels between scientists and the public still resemble those of 1989, which were specifically one-way with a narrative controlled by a select few media outlets.

Modern SETI researchers must consider this modern paradigm for consumption of news by the public, using social media and other non-traditional outlets, when planning and executing searches for extraterrestrial intelligence.  We propose additions to the post detection protocols as they pertain to the use of the Internet and social media, as well as pre-search protocols.  It is our belief that such protocols are necessary if there is to be a well-informed, sane global conversation amongst the world's citizens following the discovery of intelligent life beyond the Earth.

**Keywords: SETI, Astrobiology, protocols, science communication**

## Introduction

A critical concern for scientists pursuing the Search for Extraterrestrial Intelligence (SETI) is the reaction of the world to the knowledge that humans are not the only technological civilisation in the Universe.  The "culture shock" that results from such an epoch-making discovery has been the subject of intense discussion (cf Almár and Tarter 2011, Almár 2011, Elliott and Baxter 2012).

We can only guess at human civilisation's response to discovering ETI, but we are afforded clues by some events in recent history. A famous example is the "War of the Worlds" radio broadcast, which purported to describe a true invasion of the Earth by Martians (Cantril 1940). Reactions to the news were mixed, with many parts of the world rightly dismissing the story as fictional. In parts of South America, later broadcasts were taken at their word - in Ecuador, the resulting riots would claim the lives of up to twenty people (Gosling 2009).

This "culture shock" concern, along with many other issues such as censorship, data protection and the personal safety of the scientists involved, led the IAA to draw up non-binding post-detection protocols (IAA,1989). These protocols were to act as a guide from the instant of potential discovery, through independent verification to the public announcement of the detection.

When the protocols were first drafted in 1989, the consumption of news media followed quite traditional archetypes. The principal outlets of credible news stories were print newspapers and magazines, radio and television. Of the above, radio and television were the most responsive, able to disrupt scheduled broadcasts to bring breaking news, with CNN having just begun 24 hour news broadcasts in 1980[1]. The Internet remained a niche territory, yet to be exploited by mainstream news outlets. The 1989 protocols say very little on the subject of announcing a detection to the public, implicitly assuming that established communication channels would be rigid and one-way, as they were at the time of writing.

In the modern world, print newspapers, radio and television remain, but are now accompanied by blogs, tweets and other social media. At the time of going to press, the public consumes news in a variety of new formats, from 140 character tweets to Facebook status updates, Reddit forum threads and YouTube videos. Information traverses social networks at astonishing speeds, to the extent that natural disasters such as earthquakes can have their epicentres identified in real time via geotagged tweets from the region (Sakaki et al 2010).

While breaking news stories often begin in the domain of "conventional" media outlets through their web domain, they can equally begin as unsubstantiated rumour or gossip, amplified exponentially by the echo chambers of social media. Even when a news story is credible, well researched and broken by a trusted source, the subsequent re-telling of the story through other news outlets can warp and distort the narrative until it is unrecognisable to those at the story's origin. This is a common outcome of science press releases (e.g. Sumner et al 2014), and it seems clear that a ETI-detection story would fall victim to the same consequences.

The IAA protocols were revised and streamlined in 2010[2] (Baxter and Elliott 2012), but there remains no specific guidance for scientists' communication with the general public using modern methods of news dissemination. SETI scientists must be prepared to not simply announce a detection via press release, but to be a trusted voice in the global conversation that will begin after the initial announcement. This will require both pre-search and post-detection protocols to be implemented, which we describe below.

---

[1] https://www.youtube.com/watch?v=rWhgKuKvvPE (CNN's first hour of broadcast, June 1st 1980)
[2] http://avsport.org/IAA/protocols_rev2010.pdf

**Pre-Search Activities**

Before any search is attempted, the researchers should be prepared for what is likely to be an unprecedented media onslaught. Scientists should develop their public outreach and science communication skills as a matter of routine, but we can think of fewer circumstances where such training could be more crucial. There are a multitude of classes available to university researchers as part of continuing professional development (CPD) training programmes, as well as other third party sources.

It is common practice for science collaborations to write status updates of their research, including blogs, tweets and social media updates. This must be initiated at the pre-search stage - this content is an essential opportunity for the researchers to give a clear description of the experiment's objectives, its limitations, and their criteria for what constitutes a discovery. The team should also prepare for post-detection by agreeing on how various resources should be pooled, distributing responsibilities for various forms of public communication, and the drafting of post-detection statements. The protocols for the team's post-detection activities should be made public pre-search. This content will be studied in fine detail by a variety of agents during the post-detection phase (as well as during unintentional leaks or inaccurate rumours).

This extreme attention will also be focused on the scientists themselves. We all leave a digital footprint behind us, and the academic world depends heavily on electronic communications such as email, teleconferencing, and more recently social networking (at the time of writing, several Facebook groups exist for professional astronomers, at both general and field-specific levels, with membership numbers in the thousands). The multiplicity of online accounts for a variety of uses, and ever changing security/privacy controls leaves scientists vulnerable.

There may be many reasons for individuals to target scientists involved in the detection of intelligent life. The "Gamergate" affair[3] has shown society the extent to which an individual's life can be placed under attack by malicious agents. The publication of personal details such as home addresses online (known colloquially as "doxxing") can be particularly damaging, especially when such publication is accompanied by an incitement to violence.

SETI scientists should therefore assess their online presences. Appropriately secure passwords should be employed whenever possible, and privacy controls should be set to maximum. Personal websites should also be carefully inspected for potentially dangerous personal information. Email should be given special consideration, and scientists should take care with official email correspondence, as many publicly funded researchers are liable to

---

[3] https://en.wikipedia.org/wiki/Gamergate_controversy

Freedom of Information (FOI) requests, not to mention hacking and leaks.  In defence against the latter, it may be worthwhile to consider suitable encryption strategies. Misinterpretation of such correspondence can place the credibility of the scientists and their experiments in jeopardy, as was the case during the "ClimateGate" email scandal at the University of East Anglia (cf UK Government 2010).

**Tentative Detection, and the First Announcement**

The initial detection of what is deemed to be intelligence will likely be tentative.  A good case study is the famous "Wow!" signal, a powerful narrowband detection at around 1420 MHz that remains unverified (see Gray and Marvel 2001).  Ideally, experiments define pre-search what their criteria are for a confirmed detection, what information they will release immediately, and what verification steps are taken before the information is released.  This should be published online pre-search as described in the previous section.

As already described in the IAA protocols, the team should attempt to have their tentative discovery independently verified.  The verification process may take several days, and it may even take months or years to achieve a high level of confidence in the signal.  An illuminating case is the false alarm signal detected in 1997 by Phoenix at the Green Bank Telescope (Shostak and Oliver, 2000).  A major newspaper approached the team during the confirmation process, before it was understood to be a signal originating from the SOHO satellite observatory.  The fact that the newspaper "got wind" of a SETI detection illustrates the need for great care during this tentative phase.  As Shostak and Oliver note, it is very easy for an unscrupulous journalist to publish a "Scientists detect intelligent life" story before confirmation, as it constitutes a win-win scenario: either the journalist is first to the biggest story of their career, or they publish a story that gathers a great deal of attention and is eventually retracted with little harm to their career.

The best solution to this issue is to break the news oneself.  Having established communication channels pre-search, the team should announce their tentative detection forthwith. The announcement should coincide with a traditional press release, and the submission of a paper for peer review (with a preprint published freely online).  The handling of this stage is most crucial - the team must be as certain as possible that they cannot self-refute the detection before publishing, but undue delays can result in the unpleasant scenario described above.  The team should be clear in their published material that the signal remains unverified, and that until it is verified, it must be assumed to be caused by astronomical or human-made phenomena until proven otherwise.

Once published, the detection can then be compared by independent teams against the pre-specified criteria, and the scientists will have direct control over what content is presented to the public, including the verification process (see e.g. The BICEP2 announcement of a detection of primordial gravitational waves in the cosmic microwave background, Ade et al 2014).  This will provide foundations for a second official press statement to be made once the detection is either confirmed or shown to be spurious.  Ideally, statements for both eventualities should be drafted well in advance and agreed upon by all team members.

Data pertaining to the detection should be made available on a separate server, with several mirror sites to avoid traffic overload.  This has the advantage of allowing independent confirmation by as many groups as are interested, which is likely to be the vast majority of the SETI community, as well as citizen science programmes such as the Zooniverse.

This could even be a shared resource amongst many SETI teams, as was recommended by Shostak and Oliver in their Immediate Reaction Plan.  Such a system would be an ideal antidote against false alarms and hoaxes, undercutting the tendency to observe SETI as benighted by "government cover-ups", and generally reducing the available fodder for conspiracy theorists.  It would also provide traditional media with a host of contacts in the discovery team and the SETI community to approach for comment.  The advent of cloud computing technology such as Amazon Web Services or Microsoft Azure allows for significantly improved data storage and retrieval services, often far exceeding the capabilities of most academic institutions. Hosting data on these services would allow an extra level of redundancy, reliability, and most likely security.

This is the beginning of what we dub "the global conversation".  It is at this instant that the team must be ready to interact with the public.  Even if the signal is erroneous (see next section), the following days and weeks are likely to be the busiest, most intense of the scientists' careers.

**If the Signal is Not Confirmed**

If the detection cannot be independently verified, or is convincingly refuted, the scientists must issue a clear statement to this effect.  They must be candid about the likelihood of this signal being intelligent in origin, and if necessary retract any claims that might have been made that are no longer supported by the evidence.

The apparent detection of "faster-than-light neutrinos" by the OPERA instrument in 2011 is a useful example. An apparently unphysical experimental result, violating Einstein's theory of special relativity, was presented to the public as precisely that - anomalous, and presumably incorrect.  The authors published the data primarily to help them detect flaws in their experiment, which were duly discovered, and later tests showed the phenomenon was erroneous (OPERA Collaboration, 2012).

While it might be argued that such an approach can be damaging to a scientist's credibility, that damage comes from the handling of the story, not the story itself.  If the team has clear detection criteria, and the signal does not satisfy those detection criteria, and the narrative of growing or decreasing confidence in the detection is communicated clearly to the public, there should be no danger in going public with a signal that turns out to be a satellite, or indeed a microwave oven (Petroff et al 2015).

An excellent example of well modulated discussion of open source SETI-relevant science is the anomalous Kepler star KIC 8462852 (Boyajian et al 2016). The extremely odd transit light curves obtained from this object still challenges astronomers for a full explanation, but the current best theory involves a passing of a large cloud of comets between the star and the Earth. A more extreme explanation involves artificial megastructures, which have been shown to be visible in transit data (Forgan 2013, Korpela 2015, Wright et al 2016).

The astronomers involved in this discovery were not at liberty to keep KIC 8462852 private, due to the public nature of the Kepler data archive. Instead, the original publication of the data made clear that the SETI explanation was valid, but should be assigned a low prior probability. The media coverage, while hysterical in places, was largely measured thanks to careful guiding of the narrative by the astronomy community and interested stakeholders in the media and other institutions.

**If the Signal is Confirmed**

If the signal is indeed independently verified, then the team should announce this using the channels described previously. At this stage, they should be deeply involved in the global conversation, enriching the quality of information flowing within, and engaging fellow scientists and science communicators to contribute to the conversation. It is likely that this conversation will last for years, perhaps decades, and the team must commit to remaining part of the discussion for as long as humanly possible.

Participation within the global conversation will require interactions across as many platforms as possible. Limiting discussion to say Twitter instantly rules out demographics that do not tweet. It is the duty of scientists to engage with all citizens to the best of their ability, and a single soap box will not be enough.

Scientists must also take due care to enter the global conversation through popular outlets outside of their cultural domain. For example, Western researchers should consider social networks such as Qzone in China when attempting to disseminate their narrative. Working with colleagues and collaborators across national borders is extremely helpful in this regard.

**If Other Teams or Individuals Make A Detection**

It should be clear that even if a SETI team does not make a given discovery, they should still participate in the global conversation. As advocates for clear scientific thinking and analysis, any SETI team should be prepared to assist in confirming or refuting a detection, and engaging with the public through their available communication channels. There will be a great need for informed, sane voices, and SETI teams are best qualified to fulfil this role.

**Summary of the Proposed Protocols**

Pre-Search Protocols

* SETI scientists should gauge how the public consumes news *at the beginning of the search*. The team should maintain at minimum a blog or equivalent stream of medium- to long-form news and updates, as well as microblogging accounts such as Twitter and video publishing platforms like YouTube and Vimeo.  These are only examples of news consumption at the time of writing, and are likely to change with time. The research team must select communication channels that suit contemporary news consumption habits.

* Initial communications with the public should include detailed descriptions of the experiment setup and progress, as well as criteria for tentative and confirmed detections.  This content should also detail the protocols that the scientists will follow, both in terms of the experiment and how the experiment will be communicated to the public.

* The scientists should undergo science communication and media training, and practise these skills wherever possible.

* Throughout, the scientists must establish competence and trustworthiness prior to any detection, by a) maintaining a professional social media profile, b) documenting the experiments as they are constructed and executed, via academic papers, conventional press, and social media, and c) developing strong relationships with journalists and decision makers.

* The scientists should review their online digital footprint, and ensure that their privacy controls on social media and other websites are appropriate.  Extra efforts should be taken to avoid any personal information not relating to their work becoming publicly available, to guard against attacks from malicious agents who are opposed to changes in their worldview (cf "Gamergate" and "doxxing").  Publicly funded research teams should also be aware that their email correspondence may be published under Freedom of Information (FOI) protocols.

Post-Detection Protocols

*  In the event of a tentative detection, SETI scientists should submit their findings for a peer reviewed publication, while simultaneously publishing their candidate signal, and all accompanying data in full, using suitably robust web and cloud-based servers.  This publication should be accompanied by clear, concise statements regarding their confidence in the detection, and an acknowledgement that until proved otherwise, the tentative signal should be assumed to be caused by natural or human-made phenomena.  SETI scientists must then become extremely active in the global conversation regarding the (potential) discovery of intelligent life.  They should attempt to engage other scientists, scholars of all traditions, and science communicators to give unbiased factual accounts of the discovery, and to debunk erroneous statements.  This will involve heavy use of the social media tools above.

* If the detection cannot be confirmed, then the team must publish a statement clearly stating that the signal cannot be confirmed to be of ETI origin. If necessary, retractions should be published for any claims that cannot be held up to the evidence.

* Once the detection is confirmed, SETI scientists must commit to continued participation in the global conversation, for as long as is humanly possible.

## Acknowledgements

This work was originally presented at the 2015 meeting of the UK SETI Research Network at Leeds Beckett University.   The authors are grateful to the UKSRN members for their insightful comments.